\documentclass[11pt]{article}

\usepackage{amsfonts,amssymb,eucal}
\usepackage{amsthm}
 \usepackage{amsmath}
\usepackage{amscd}
 \usepackage{latexsym} 
 \newcommand{\got}[1]{{\mathfrak{#1}}}% gothic with mbox for  mathematic
\newcommand{\db}[1]{{\mathbb{#1}}}% double
\numberwithin{equation}{section}

\newcommand{\R}{\ensuremath{\mathbb{R}}}
\newcommand{\C}{\ensuremath{\mathbb{C}}}

\begin{document}

\title{Finite-level systems, Hermitian operators, isometries, and a novel parameterization of Stiefel and Grassmann manifolds}
\author{Petre Di\c t\u a\\
Institute of Physics and Nuclear Engineering,\\
P.O. Box MG6, Bucharest, Romania\\ email: dita@zeus.theory.nipne.ro}

\maketitle

\begin{abstract}

In this paper we obtain a  description of the Hermitian operators acting on the Hilbert space $\C^n$, description which  gives a complete solution to the over parameterization problem. More precisely we provide an explicit  parameterization of arbitrary $n$-dimensional operators,  operators that may be considered either as Hamiltonians, or density matrices for  finite-level quantum systems.  It is shown that the spectral multiplicities are encoded in a flag unitary matrix obtained as an ordered product of special unitary matrices, each one generated by a complex $n-k$-dimensional unit vector, $k=0,1,\dots,n-2$. As a byproduct, an  alternative and simple parameterization of Stiefel and Grassmann manifolds is obtained.
\end{abstract}
%%%%%%%%%%%%%%%%%%%%%%%%%%%%%%%%%%%%
\section{Introduction}
%%%%%%%%%%%%%%%%%%%%%%%%%%%%%%%%%%%%

There is a considerable interest in a simple description of density matrices that have a wide variety of applications, particularly in quantum information theory, and many efforts were devoted in describing them. However the problem of over parameterization is still open \cite{Zy},  for an arbitrary $n$. We solve this problem by  providing an explicit parameterization of eigenvalues, as well as of the unitary matrices that diagonalize  arbitrary finite-dimensional Hermitian operators. 
 Quite expectedly, such a description is closely related to the description of various homogeneous manifolds, alluded to the title.On the other hand the
  Stiefel, or Grassmann  manifolds arise  in many problems from 
  different other domains  such as  encryption, coherent states, geometric phases, signal processing, geometric integration on homogeneous manifolds,  numerical  linear algebra algorithms, and many others. %These manifolds became a first hand tool in solving many problems in applied mathematics, mathematical and theoretical physics, and the diversity of the problems necessitates to keep improving it.
 In such problems, the
``states of interest''  are, in  general, elements of some homogeneous space
\begin{eqnarray}{ X\cong G/K}\end{eqnarray}
where $G$ is a Lie group, and $K$ is a closed  subgroup of $G$. Further in  problems arising  in engineering, physics, quantum information theory, one needs a concrete realization of these manifolds in a form able to be stored into a computer. Although, the geometrical description of  Grassmann and Stiefel manifolds is available in many books, see for example \cite{G}, \cite{KN}, the available parameterizations of the above manifolds are not the most convenient in some concrete applications, e.g. \cite{EAS},\,\cite{ZT} .

One aim of the paper is to obtain a satisfactory description of Hermitian operators that appear in the study of finite-level systems. By satisfactory, we mean a complete as possible description of equivalence classes of Hermitian operators, i.e. a full and complete parameterization of the orbits generated by these operators. We recall that two Hermitian operators are in the same orbit if their spectra coincide, or, equivalently, if their characteristic polynomials are identical. Our own interest in studying such  problems was originally awakened by  a query raised by a colleague of mine \cite{S} who is interested in a complete description of the solutions of the
 quadratic operator equation
\begin{eqnarray}{\rho^2-2 p\,\rho +(p^2-q^2)\,I_n=0,\quad p,q \in {\R},\,\, {\rm with} \,\, p^2-q^2 \ge 0 \label{qq}}\end{eqnarray}
where, in the following, $I_n$ denotes the unit $n$-dimensional matrix, and $\rho$ is a Hermitian operator. When $\rho$ is a density matrix,  the  equation 
(\ref{qq})   is the simplest generalization of the pure state condition $\rho^2=\rho$. We note that  Werner states \cite{We}, and Horodecki states \cite{Ho} satisfy such an equation.
 
By itself, an equation  as (\ref{qq}) has nothing special. According to Cayley-Hamilton theorem, any finite $n$-dimensional matrix satisfies an $n$-degree polynomial equation, the characteristic polynomial.  As we will see in the following, equation (\ref{qq}) describes a Hermitian operator whose spectrum has a maximum degeneracy. If the  multiplicities of eigenvalues  are denoted by $k$ and $n-k$, respectively,   the spectral decomposition of   $\rho$ is
\begin{eqnarray}{\rho =\lambda_1\,P +\lambda_2(I_n-P)\label{pro}}
\end{eqnarray}
where $P$ is the projection onto the  $k$-dimensional subspace. Thus, a description of the  projection $P$  on  ${\C}^n$ is equivalent to a description of the  Grassmann manifold $Gr(k,n)$.

Formally, the states of an $n$-level quantum system are described by density matrices $\rho$ that are positive, Hermitian, nuclear operators whose trace is normalized to unity
\begin{eqnarray}{\rho=\{\rho\ge 0,\,\, \rho=\rho^*,\,\,Tr\,\rho=1\}}\end{eqnarray}
where $^*$ denotes adjoint, i.e. the complex conjugated transpose.

If we denote by $D=(\lambda_1,\dots\lambda_n)$ the diagonal matrix of $ \rho$ eigenvalues,  elementary facts from the spectral theory of self-adjoint operators tell us that there exists a unitary operator $U\in U(n)$, where $U(n)\in End\,{\db C}^n,$   generated by the eigenvectors of  $ \rho$   such that
\begin{eqnarray}{\rho=U\,D\,U^* \label{He}}\end{eqnarray}
with $\sum_1^n\lambda_i=1$. In the  generic case, a Hermitian $n\times n$ matrix $H$ is parameterized by   $n^2$ real parameters, number which coincides with the number of parameters entering the parameterization of an arbitrary unitary matrix $U\in U(n)$. Since the eigenvalues $\lambda_i$ are also independent, we infer from (\ref{He})
 that there are some constraints upon the form of $U$, i.e. the number of free parameters entering $U$ is less than $n^2$.  It is well known that the constraints coming from the Hermitian character of an operator $H$ may be translated to the request that its eigenvectors are defined up to an overall arbitrary phase. We choose the phases such that the first entry of each eigenvector is a  non-negative number. This means that the first row entries  of $U$ are  non-negative, i.e. $U$ is parameterized by $n(n-1)$ real parameters, and in this way we conclude that a generic Hermitian matrix is parameterized by $n^2$ real parameters, as it should be. In fact this approach will be used everywhere in the paper: we start with a parameterization of $U(n)$ and restrict to the appropriate subset of coordinates to describe the corresponding manifold.

We made this digression because, especially in the physical literature, $U$, entering equation (\ref{He}),
is considered  an element of $SU(n-1)$, that evidently leads to an over parameterization. Thus when we have to do some symbolical, or even numerical calculations, as in \cite{Zy} or \cite{Sl}, we have to be more careful.  We shall see later that such a $U$ is a matrix realization of the flag manifold
\begin{eqnarray}{X \cong U(n)/U(1)^n}\end{eqnarray}
where $U(1)^n$ denotes the torus subgroup of $U(n)$. By
 obtaining a full and explicit parameterization of $X$, we obtain, via formula (\ref{He}), a parameterization of all finite-dimensional Hermitian operators whose spectra are simple.

Such a construction addresses an old fundamental question in the theory of measurement \cite{Wi},\cite{La}, namely if it is possible to measure experimentally the 'variables' corresponding to an arbitrary Hermitian operator. Thus the first question to be solved is the finding of the 'variables' entering an Hermitian operator. After that, the answer is simple:  there does exist an experimental embodiment for every Hermitian operator in finite-dimensional Hilbert space; see \cite{RZBB} for details concerning its  realization.

If $\rho$ is a density matrix, then   $\rho$ in equation (\ref{He})  is parameterized by $n^2-1$ parameters because of the trace condition.
At the other extreme, there is the case when $H$ is an one-dimensional projection, and then $U$ entering (\ref{He}) is a matrix realization of the coset
\begin{eqnarray}{X \cong \frac{U(n)}{U(1)\times U(n-1)}\cong Gr(1,n)}\cong {\bf C}P^{n-1}\nonumber\end{eqnarray}
where $G(1,n)$ denotes the simplest Grassmannian. Between these two extreme cases lie all the other spectral types. For example, the solutions  of equation (\ref{qq}) describe the most degenerate spectrum, which means that $\rho$ has only two distinct non-zero eigenvalues. By  description,  we will understand the parameterization of the set of  all unitary diagonalizing operators $U$ in terms of a subset of parameters entering the parameterization of the group $U(n)$, for a generic situation.  In other words we are looking for a matrix realization  of the relation  (\ref{pro}), i.e. of the projection $P$, and this is the place where the Grassmann manifolds enter the play; and we expect that in this case, stronger conditions on the form of $U$ will be in force. In fact, in this paper, we provide a unified method for treating   all the spectral types of  Hermitian operators, by properly taking into account their spectral multiplicities.

 This problem is closely related to  the description of  isometries between the Hilbert spaces $\db C^k$ and  ${\db C}^n$,\,$1\leq k\leq n$. The isometries  are operators  generated by  $n\times k$ or $k\times n$  matrices whose  columns, and respectively rows, are   orthogonal, and in the following we show that there is a close relationship between these isometries and different matrix realizations of the  coset spaces generated as in (1.1). The necessity of working with matrices that have orthogonal column vectors, or row vectors, became evident in the last years; see e.g. \cite{EAS}, \cite{FGP}.

The mathematical background necessary for obtaining such results  are elementary facts from the spectral theory of self-adjoint operators and the  theory of contraction operators, and a trivial lemma that we state here for the case of the $n$-dimensional unitary group $G=U(n)$; for the general case see \cite{G}.
\newtheorem{Th} {Lemma}
\begin{Th}
{\it 
The coset relation

\begin{eqnarray}{X\cong U(n)/K\label{Le}}\end{eqnarray}
where $K \subset U(n)$ is a subgroup, can be written as a matrix relation in the following form
\begin{eqnarray}{M_n=A_n\,B_K}\label{Lee}\end{eqnarray}
where $M_n\in U(n)$ is an arbitrary $n\times n$ unitary matrix, $A_n\in U(n)$
is a unitary matrix parameterized  by a point of the coset $X$, and $B_K\in K$ is an arbitrary element of the subgroup $K$ viewed as an element embedded in $U(n)$}.
\end{Th}

By using it we obtain new parameterizations of flag, Stiefel and Grassmann manifolds. The paper is a sequel of our results concerning the factorization of unitary matrices \cite{Di2} in terms of $n$ complex  vectors $v_i\in S^{2i-1},\,\, i=1,\dots,n$, where $S^{k}$ is the $k$-dimensional sphere in ${\db C}^n$.
We denote by $M(n,k)$ the set of all $n\times k$ complex matrices over ${\db C}^n$, and the main mathematical result of the paper is: 
\vskip2mm

{\bf Main Theorem.}\,\,{\it 
 Let $\mathfrak{C}\in M(n,k)$ be an  $n\times k$ complex matrix  that generates an isometry, i.e. $\got{C}: \,\, {\db C}^k \rightarrow {\db C}^n$, \,\, $\got{C}^*\,\got{C}=I_k$. Then, the matrix representation of the coset generated by the point $\got{C}$ is realized in terms of a unitary matrix $A(n,k)$ that diagonalizes the projection $D_{T^*}$, where $D_{T^*}$ is the defect operator associated to the isometry $\got{C}$,
  under the form  \begin{eqnarray}{I_n-D_{T^*}=A(n,k)\left(\begin{array}{cc}
I_k&0\\
0&0_{n-k}
\end{array}\right)A(n,k)^*=\sum_{i=1}^{k}\,c_i\cdot c_i^*=\got{C\,C}^* }\label{Mth}\end{eqnarray}  where $c_i$, $i=1,\dots,k$, are the (orthogonal) column vectors of $\got{C}$. }  
\vskip2mm
All the other results discussed in this work are a consequence of the above theorem.

The organization of paper is as follows. In section 2 we show the close relationship between the isometries  $ \mathfrak{C}\in M(n,k)  $ and matrix realizations of the coset spaces. In section 3 we reformulate our  results \cite{Di2} in a more convenient form for the present applications, and we find the generic parameterization of $n$-dimensional Hermitian operators.  In section 4 we obtain our matrix  parameterizations of Stiefel and Grassmann manifolds. The paper ends with Conclusion.

%%%%%%%%%%%%%%%%%%%%%%%%%%%%%%%%%%%%%%%%%%%%%%%%
\section{Isometries}
%%%%%%%%%%%%%%%%%%%%%%%%%%%%%%%%%%%%%%%%%%%%%%%%

 In this section we show  how the isometries  $ \mathfrak{C}\in M(n,k)  $ can be used for the parameterization of various interesting manifolds. The main idea is that the  isometries   generated by $k$ rows, or columns of an arbitrary $n\times n$  matrix, $1\leq k\leq n$, allow  to define  projection operators whose spectral decomposition provides the necessary tool in finding matrix representations of various interesting  manifolds.    For what follows,  we need a few elementary notions from contraction operator theory; we use this theory since it has a powerful functional calculus that help us in doing explicit calculations.

An operator $T$ applying the Hilbert space $\cal{H}$ in the Hilbert space
$\cal{H}'$ is a contraction if for any $v\in
{\cal{H}}$,~~$||T\,v||_{{\cal{H'}}}\leq ||v||_{{\cal{H}}}$, i.e. $||T||\leq
1$, where $||T||$ denotes the norm of $T$ \cite{SF}. For any contraction we have $T^*\,T\leq I_{{\cal{H}}}$ and $T\,T^*\leq I_{{\cal{H'}}}$,  where $T^*$ denotes the adjoint, that is defined by the relation  $(Tv,v')=(v,T^* v')$, \, $v\in {\cal{H}}$,\,$v'\in {\cal{H'}}$, and $(\cdot,\cdot)$ is the usual inner product in $\cal {H},\, {\rm or,\,\ respectively,}\,\cal {H}'$. To any contraction $T$ one associates two defect operators 
 by  the relations

\begin{eqnarray} {D_T=(I_{{\cal{H}}}-T^*\,T)^{1/2},\quad
D_{T^*}=(I_{{\cal{H'}}}-T\,T^*)^{1/2}\label{Def} }\end{eqnarray}
that  are Hermitian operators  in
${{\cal{H}}}$ and ${{\cal{H}'}}$, respectively. They have the property 
\begin{eqnarray}{  T\,D_{T}=D_{T^*}\,T,\quad T^*\,D_{T^*}=D_{T}\,T^*}\label{com}\end{eqnarray}
Finite-dimensional  contractions can be seen as  being generated by $n\times k$, or  $k\times n$ matrices, and for definiteness we consider the first case, i.e. $T$ has $n$ rows and $k$ columns and we denote it by $\got{C}$.
In the following we are interested in  contractions of a special form,
namely the isometries between ${\db C}^k$ and  ${\db C}^n$. They are operators  $T:\, {\db C}^k \rightarrow\,{\db C}^n\,$ that  satisfy $T^* T=I_k$, \,$k=1,\dots,n$, i.e. the columns of  $T$ are {\it orthogonal} columns, and respectively, $T:\, {\db C}^n \rightarrow\,{\db C}^k\,$  that satisfy   $T\,T^*=I_k$,  when the  $k$  rows are {\it orthogonal}. With our choice, i.e.  $T$ is a $n\times k$ matrix, we have the identification ${\cal H}\equiv {\db{ C}}^k$ and ${\cal H'}\equiv {\db{ C}}^n$.

 For an isometry $\got{C}$ generated by $k$ columns, the relations  (\ref{Def}) give  $D_T=D_{\got{C}}=0$. However, in the following we preserve the  notation for the defect operator $D_T$, to be implied that it is generated by a definite contraction,  and from the first relation (\ref{com}) we deduce that 
 \begin{eqnarray}{D_{T^*}\,T= T\,D_T=0= D_{T^*}\,\got{C}=
 \lambda\,\got{C}}\end{eqnarray}
i.e. all columns of ${\got{C}}$ are in the kernel of $D_{T^*}$. In other words, the $k$ column vectors are the eigenvectors that correspond to the eigenvalue $\lambda = 0$, and these eigenvectors are orthogonal. When  $D_T=0$, the other defect operator $D_{T^*}$ is a projection, i.e. a self-adjoint operator which satisfies $D_{T^*}=D_{T^*}^2$, and we infer that rank $D_{T^*}= n-k$; thus $I_n-D_{T^*}$ projects onto the eigenspace corresponding to the  eigenvalue $\lambda =1$.  Then, from the preceding relation, we get
 \begin{eqnarray}{(I_n-D_{T^*})\got{C}= I_n\,\got{C}=\got{ C}=\lambda\got{ C}}\end{eqnarray}
i.e. the $k$ column vectors of $\got{ C}$ are the orthogonal eigenvectors of  $I_n-D_{T^*}$ corresponding to the $k$ eigenvalues $\lambda=1$, or in other words, rank\,($I_n-D_{T^*})$ = $k$.

For us the interesting objects are the unitary matrices that diagonalize the projections $D_{T^*}$ and $I_n-D_{T^*}$.
 Let $A(n,k)$ be the unitary matrix  which diagonalizes the projection $D_{T^*}, $ then
 \begin{eqnarray}{ A(n,k)^*D_{T^*}A(n,k)=\left(\begin{array}{cc}
0_k&0\\
0&I_{n-k}
\end{array}\right)}\end{eqnarray}
and from it we obtain a matrix representation of the projection $D_{T^*}$ under the form 
 \begin{eqnarray}{D_{T^*}=A(n,k)\left(\begin{array}{cc}
0_k&0\\
0&I_{n-k}
\end{array}\right) A(n,k)^*} \label{Eq}\end{eqnarray}
But  we also have
 \begin{eqnarray}{ A(n,k)^*(I_n-D_{T^*})A(n,k)=\left(\begin{array}{cc}
I_k&0\\
0&0_{n-k}
\end{array}\right)}\end{eqnarray}
relation which can be written as
\begin{eqnarray}{I_n-D_{T^*}=A(n,k)\left(\begin{array}{cc}
I_k&0\\
0&0_{n-k}
\end{array}\right)A(n,k)^*=\sum_{i=1}^{k}\,c_i\cdot c_i^* =\got{C}\,\got{C}^* }\label{Eqq}\end{eqnarray} where $c_i,\, i=1,\dots,k$, are the column vectors of the isometry $\got{C}$.
Equations  (\ref{Eq}) and  (\ref{Eqq}) provide   matrix representations for both the projections on the $n-k$-, and respectively $k$-dimensional, subspaces of ${\db C}^n$. In our applications $A(n,k)$ will be an  $n\times n$ unitary matrix generated by a {\it coset} as in {\bf Lemma 1},  coset, which  at its turn, is parametrized by the {\it point} $\got{ C}$. 
\vskip2mm

The above representation does not provide a full explicit form for the matrix $A(n,k)$. We know only that its  first $k$ columns coincide with the $\got{C}$ columns. Thus an important problem is the completion of  the  matrix $A(n,k)$ with $n-k$ columns {\it without introducing new parameters}, i.e. the next $n-k$ columns  must  be determined by the first $k$ columns. For doing that we need a parameterization of the first $k$ columns and the most convenient one is to introduce generalized spherical coordinates. In the next section we show that the most important case is $k=1$, the other cases being a direct consequence of it, which leads to a factorization of unitary matrices. In the same time,  we reformulate our results from \cite{Di2}, concerning  factorization of unitary matrices, that will provide  the necessary tools for obtaining matrix realizations for symmetric manifolds. 

%%%%%%%%%%%%%%%%%%%%%%%%%%%%%%%%%%%%%%%%%%%%%%%%%%
\section{Factorization of unitary matrices}
%%%%%%%%%%%%%%%%%%%%%%%%%%%%%%%%%%%%%%%%%%%%%%%%%%

The idea behind such a factorization comes from the following sequence
\begin{eqnarray}{U(n)\cong \frac{U(n)}{U(n-1)}\times\frac{U(n-1)}{U(n-2)}\times\dots\times\frac{U(2)}{U(1)}\times U(1)\nonumber}\end{eqnarray}
\begin{eqnarray}{\cong S^{2n-1}\times S^{2n-3}\times\dots\times S^3\times  S^1~}
\end{eqnarray}
sequence that shows that each factor can be parameterized by an arbitrary point on the corresponding complex sphere. 

 The factorization comes from the relation (3.1) which we write as
\begin{eqnarray}{S^{2n-1}\cong \,U(n)/U(n-1)}\end{eqnarray}
Similarly to relation (\ref{Lee}), we  rewrite (3.2)  as
\begin{eqnarray}{ M_n=B_n^0\,M_{n-1}^1}\end{eqnarray}
where  $ M_n\in U(n)$ is an arbitrary $n\times n$ matrix, $B_n^0\in U(n)$ is a special matrix generated by  one of its column vectors, for example by its first column vector, which at its turn is parameterized by a point $z\in S^{2n-1}$,  and
\begin{eqnarray}{M_{n-1}^1=\left(\begin{array}{cc}
1&0\\
0&M_{n-1}\end{array}\right)}\end{eqnarray}
where $M_{n-1}\in U(n-1)$ is an arbitrary $(n-1)\times(n-1)$ matrix. For further details see \cite{Di2}.
By iteration of (3.4), we arrive at the  form
\begin{eqnarray}{ M_n=B_n^0\cdot B_{n-1}^1\dots B_1^{n-1}}\end{eqnarray}
 where each $B_{n-k}^k$ is generated by a column vector, and here is the point where an explicit parametrization enters.
First we remind that a standard parameterization of   $U(n)$ is given in terms of $ n(n-1)/2$ angles, $\theta_i\in [0,\pi/2],\,\, i=1,\dots,n(n-1)/2, $ and  $ n(n+1)/2$ phases,
$\varphi_i \in [0,2\,\pi),\,\, i=1,\dots,n(n+1)/2$, see e.g.\,\,\cite{Di1},\,\, \cite{M}.

The matrix realization of formula (3.5), which is the main result in \cite{Di2}, is given by 

\newtheorem{Le} {Theorem}

\begin{Le}
{\it Any  element $M_n\in U(n)$ can be factored into an ordered product of $n$ matrices of the following form}
\begin{eqnarray}{ M_n=B_n^0\cdot B_{n-1}^1\dots B_1^{n-1}\label{mat}}\end{eqnarray}
 where 

\begin{eqnarray}{ B_{n-k}^k=\left(\begin{array}{cc}
I_k&0\\
0&B_{n-k}
\end{array} \right)}\nonumber\end{eqnarray}
{\it and $B_{n-k}\in U(n-k)$ are special 
unitary matrices,  each one  generated by a single complex  $(n-k)$-dimensional unit vector, $b_{n-k}\in S^{2(n-k)-1}$. For example $B_1^{n-1}=e^{i\varphi}$, where $\varphi$ is an arbitrary phase.}{\it

If $y_m\in S^{2m-1},\,\,\, m=1,\dots,n$, is parameterized by
\begin{eqnarray}{y_m=(e^{i\varphi_1}\cos\,\theta_1,e^{i\varphi_2}\sin\,\theta_1\cos\,\theta_2,\dots,e^{i\varphi_m}\sin\,\theta_1\dots \sin\,\theta_{m-1})^t\nonumber}\end{eqnarray}
where $t$ denotes transpose,  the $m$ columns of $B_m$ are given by}

\begin{eqnarray}
v_1=y_m=\left(\begin{array}{l}
e^{i\varphi_1}\cos\,\theta_1 \\
e^{i\varphi_2}\sin\,\theta_1\cos\,\theta_2 \\
\cdot\\
\cdot\\
\cdot\\
e^{i\varphi_m}\sin\,\theta_1\dots \sin\,\theta_{m-1}
\end{array} \right)\nonumber \end{eqnarray}
{\it and
\begin{eqnarray}{
v_{k+1}=\frac{d}{d\,\theta_k }\,\,v_1(\theta _1=\dots
=\theta _{k-1}=\pi/2),\qquad k=1,\dots,m-1}, \nonumber \end{eqnarray}
where in the above formula one calculates first the derivative and afterwords
the restriction to $\pi/2$. }\end{Le}

The essential point of the algorithm for getting (\ref{mat}) was the use of contraction operator theory to construct {\it all} the eigenvectors of a self-adjoint projector, generated by the first column vector of $B_n^0$, eigenvectors that are easy provided by the above theorem, if the first vector generating them is parameterized in spherical coordinates.

Taking into account that by  multiplication  at left of an arbitrary matrix from $U(n)$     by a diagonal phase matrix $d_n=(e^{i\varphi_1},\dots,e^{i\varphi_n})$, the first  row is multiplied by $e^{i\varphi_1}$, the second by  $e^{i\varphi_2}$, etc., and the last one by $e^{i\varphi_n}$ we can write
 $B_n^0=d_n\, \tilde{B_n^0}$ where the first column of $\tilde{B}_n^0\in SO(n)$ has non-negative entries.

For our aims  we need an explicit parameterization of $B_{n-k}$, $k=0,\dots,n-1$, and we choose the $n$  generating vectors as follows
\begin{eqnarray}{\begin{array}{l}
y_n=(e^{i\alpha_1}\cos\,a_1,e^{i\alpha_2}\,\sin\,a_1\,\cos\,a_2,\dots,,e^{i\alpha_n}\,\sin\,a_1\dots \sin\,a_{n-1})^t\\
\\
y_{n-1}=(e^{i\beta_1}\cos\,b_1,e^{i\beta_2}\sin\,b_1\,\cos\,b_2,\dots,e^{i\beta_{n-1}}\sin\,b_1\dots \sin\,b_{n-2})^t\\
....................................................................................................\\
y_2=(e^{i\psi_1}\cos\,z_1,e^{i\psi_2}\sin\,z_1 )^t\\
\\
y_1=e^{i\omega_1}
\end{array}\label{Ve}}
\end{eqnarray}

The projection operator $I_n-D_{T^*}$ entering formula  (\ref{Eqq}) is a self-adjoint operator, and for such an operator its eigenvectors $c_i$ are defined up to an overall phase. We choose the phases such that the first entry of each eigenvector is a nonnegative number.
With the generating vectors of the form  (\ref{Ve}), the  parameterization of the matrix  (\ref{mat}), given  by {\bf Theorem 1}, is such that its  first row entries have the form
\begin{eqnarray}{\begin{array}{l}
m_{11}=e^{i\alpha_1}\cos\,a_1, \,\,\,m_{12}=-e^{i(\alpha_1+\beta_1)}\cos\,b_1\,\sin\,a_1,\\
  m_{13}=e^{i(\alpha_1+\beta_1+\gamma_1)}\cos\,c_1\,\sin\,a_1\,\sin\,b_1,\,\,\,\dots,\\
 m_{1n}=(-1)^{n-1}\,e^{i(\alpha_1+\dots +\omega_1)}\,\sin\,a_1\dots \sin\,z_1 
\end{array}}\end{eqnarray}
and if we want that these matrix elements should be nonnegative we have to take
\begin{eqnarray}{\alpha_1=0,\quad\beta_1=\gamma_1=\dots =\omega_1=\pi}\end{eqnarray}

In the following we shall use in relations (\ref{Ve}) these constraints,  and so 
we  remove the first  phase of each of  the vectors $y_k$   and change the numbering as $\alpha_i\rightarrow\alpha_{i-1},\,\,i=2,\dots,n,\,\beta_i\rightarrow\beta_{i-1},\,\,i=2,\dots,n-1,$ etc., and now each vector $y_k$ is parameterized by $2(k-1)$ parameters, $\,k=2,\dots,n$, i.e. an equal number of phases and angles. In this way the last vector disappears, such that the relation (\ref{mat}) has the following form
\begin{eqnarray}{\got{ M}_n=B_n^0\cdot B_{n-1}^1\dots B_2^{n-2}} \label{matt}\end{eqnarray}
 The last matrix is the  matrix realization   of the flag manifold
\begin{eqnarray}{Fl(n)\cong\frac{U(n)}{U(1)\times U(1)\times\dots\times U(1)}\cong\frac{U(n)}{U(1)^n} \label{mattt}}\end{eqnarray}
 it depends on $n(n-1)$ parameters and it is the most general form of a unitary matrix that diagonalizes an $n$-dimensional Hermitian operator $H$ whose all the  eigenvalues are simple. In the same time $Fl(n)$ is the 'natural' manifold on which one can define a symplectic structure, and an interesting problem would be to find its explicit form. 

Putting together the information contained in {\bf Theorem 1} and formulas  (\ref{matt}) and (\ref{mattt})  we have 
 the following
\begin{newtheorem}{Co}{Corollary}
\begin{Co}
The matrix ${\got{M}_n}$ describes the spectral decomposition of a finite-dimensional Hermitian operator $H$ whose eigenvalues are simple, i.e. they satisfy a relation as
\begin{eqnarray}{\lambda_1 >\lambda_2 > \dots
>\lambda_n}\nonumber\end{eqnarray} 
and the operator  $H$ is written in the form
\begin{eqnarray}{H=\sum_{i=1}^n\,\lambda_i\,u_i\,u_i^*}\label{Hh}\end{eqnarray}
where $u_i$, $i=1,\dots,n$, are the column vectors of the matrix  (\ref{matt}).
 If $H\geq 0$ and $Tr\,H=h \in {\db R}_+^*$,
then  $\lambda_i$, entering formula (3.12), could be parameterized as
\begin{eqnarray}{\lambda_1=h\,
 \cos^2\theta_1,\,\,\lambda_2=h\,\sin^2\theta_1\,
 \cos^2\theta_2,\,\dots,\,\,\lambda_n=h\,\sin^2\theta_1\,
\dots\,\sin^2\theta_{n-1}}%\nonumber
\end{eqnarray}
where $\theta_i\in [0,\pi/2],\,\, i=1,\dots,n-1$, are arbitrary angles.
 If $H$ is a density operator,  $h=1$ in the above relation.

If $H$ is not positive definite, let suppose that  its first $p$ eigenvalues are positive, and $n-p$ are negative. If $Tr\,H=0$, and $h =Tr\,H_{\lambda_i > 0}=- Tr\,H_{\lambda_i < 0}\in {\db R}_+^* $, then a parameterization of eigenvalues is given by 
\begin{eqnarray}{\lambda_1=h\,\cos^2\theta_1,\,\,\lambda_2=h\,\sin^2\theta_1\,\cos^2\theta_2,\,\dots,\,\,\lambda_p=h\,\sin^2\theta_1\dots \sin^2\theta_{p-1},}%\nonumber
\end{eqnarray}
\begin{eqnarray}{\lambda_{p+1}=-h\,\, \cos^2\theta_p,\,\,\dots,\,\,\lambda_n=-h\,\sin^2\theta_p\,\dots\,\sin^2\theta_{n-1}}%\nonumber
\end{eqnarray}

If  $Tr\,H=h\neq 0$,  the  parameterization of eigenvalues is given by
\begin{eqnarray}
\lambda_1=|h|\cosh^2 \theta, & \lambda_2=|h|\cosh^2 \theta\, \cos^2\theta_1,& \dots,\nonumber\end{eqnarray}
\begin{eqnarray}  \lambda_p=|h|\cosh^2 \theta\,\sin^2\theta_1\dots \sin^2\theta_{p-2}
\end{eqnarray}
\begin{eqnarray}{\lambda_{p+1}=-|h|\sinh^2 \theta,\,\,\dots,\,\,\lambda_n=-|h|\sinh^2 \theta\,\sin^2\theta_{p-1}\,\dots\,\sin^2\theta_{n-2}}%\nonumber
\end{eqnarray}
if  $h >0$, where $\theta\in {\db R}_+^*,\,\, \theta_i\in [0,\pi/2],\,\, i=1,\dots,n-2$, are arbitrary angles,
 and by a similar formula in which one interchanges  $ch^2 \theta \rightleftharpoons sh^2 \theta$, 
if $h < 0$.
\end{Co}

In this way, {\bf Corollary 1} gives a simple and explicit parameterization of all generic
finite-dimensional Hermitian
operators.
%%%%%%%%%%%%%%%%%%%%%%%%%%%%%%%%%%%%%%%%%%%%%%%%%%%%%%%%%%%%%%%%%%
\section{Matrix realizations of Stiefel and Grassmann manifolds}
%%%%%%%%%%%%%%%%%%%%%%%%%%%%%%%%%%%%%%%%%%%%%%%%%%%%%%%%%%%%%%%%%%%

Looking at relation  (\ref{mattt}) we consider that the next simpler manifold is
the Stiefel manifold that we define as  the coset space
 \begin{eqnarray}{ St(k,n)\cong\frac{ U(n)}{U(1)^k\times U(n-k)}}\cong\frac{Fl(n)}{Fl(n-k)}\label{St}\end{eqnarray}
instead of the usual definition 
 \begin{eqnarray}{ St(k,n)\cong\frac{ U(n)}{ U(n-k)}}\label{Stt}\end{eqnarray}
Both forms are  similar, only the number of parameters entering them is different.
 According to {\bf Lemma 1}  we can write the first relation (\ref{St}) as a matrix relation
\begin{eqnarray}{M_n= A(n,k)\left(\begin{array}{cc}
I_k&0\\
0&M_{n-k}
\end{array}\right)}\end{eqnarray}
where $M_n\in U(n)$,\,$M_{n-k}\in U(n-k)$\, and  $A(n,k)\in U(n)$ is the matrix realization of the Stiefel manifold $St(k,n)$  parameterized by a point represented by a $n\times k$ complex matrix, with the first row entries non-negative numbers.
Alternatively, we may consider that $M_n\in Fl(n)$,\,\,$M_{n-k}\in Fl(n-k)$.
 Looking at relation  (\ref{Eqq}) we see that $A(n,k)$ is the same object in both relations (\ref{Eqq}) and (4.3). By consequence we deduce from (\ref{matt}) that
 \begin{eqnarray}{A(n,k)=B_n^0\cdot B_{n-1}^1\dots B_{n-k+1}^{k-1}}\end{eqnarray}
is the matrix realization of the Stiefel manifold  (\ref{St}). Thus the following holds

\setcounter{Le}{1}
 
\begin{Le}{\it 
The matrix representation of the Stiefel manifold $St(k,n$) as defined by {\rm(\ref{St})} is obtained from the parameterization {\rm (\ref{matt})}
by taking zero all the parameters entering $U(n-k)$, i.e. $U(n-k)=I_{n-k}$ such that
\begin{eqnarray}{ A(n,k)= B_n^0\cdot B_{n-1}^1\dots B_{n-k+1}^{k-1}\label{matttt}}\end{eqnarray}
%\end{Le}
  By using  elementary facts from spectral theory of self-adjoint operators we find that the matrix representation of the projection $I_n-D_{T^*}$ writes as
 \begin{eqnarray}{I_n-D_{T^*}=St(k,n)= c_1\,c_1^* +\dots + c_k\,c_k^*}\end{eqnarray}
where $c_i,\,\,i=1,\dots,k$ are the first $k$ column vectors  of {\rm (\ref{matttt})}. As long as the matrix {\rm (\ref{matttt})}  is parameterized by $d_1=n^2-k-(n-k)^2 = k(2n-k-1)$ real parameters, we can choose any $k$ column vectors of {\rm (\ref{matttt})} in formula {\rm(4.6)} and we can make this choice in $\binom{n}{k}$ modes.
If the form  {\rm (\ref{Stt})} is used, then 
\begin{eqnarray}{ A(n,k)= B_n^0\cdot B_{n-1}^1\dots B_{n-k+1}^{k-1}}\end{eqnarray}
where $B_{n-p}^p, \,\, p=0,\dots,k-1$ are the first $k$ matrices entering equation  {\rm (\ref{mat})}, i.e. the complex vectors generating {\rm (4.7)} are such as in  the written form {\rm(\ref{Ve})}, comprising  $n(n+1)/2$ phases. The difference between the forms {\rm (4.5)} and {\rm (4.7)} consists in the number of real parameters entering them: {\rm (4.5)} is parameterized by $d_1$  parameters, while {\rm(4.7)} depends on $d_2=n^2-(n-k)^2=k(2n-k)$ parameters. }
\end{Le}
\vskip2mm
{\it Remark}. Similarly to $Fl(n)$, on the Stiefel manifold (\ref{St}) one can define a symplectic structure.

As a consequence of the above theorem, we have the following
%\begin{newtheorem}{Co}{Corollary}
\setcounter{Co}{1}
\begin{Co}
The matrix  {\rm(\ref{matttt})} describes the spectral decomposition of a finite-dimensional Hermitian operator $H$  that has a degenerate eigenvalue of multiplicity $k$, 
$\lambda_i=\dots =\lambda_k $, and all the other eigenvalues are simple. In this case  $H$ writes as
\begin{eqnarray}{H=\lambda_1\sum_{j=1}^k\,u_j\,u_j^* +\sum_{i=1}^{n-k}\lambda_{k+i}\,u_{i+k}\,u_{i+k}^*}\end{eqnarray} where $u_i$, $i=1,\dots,n$, are the column vectors of the matrix  {\rm(\ref{matttt})}. If $H$ is a density operator, then 
\begin{eqnarray}{\lambda_1= \cos^2\theta_1,\, \lambda_{k+1}=\sin^2\theta_1\, \cos^2\theta_2,\dots, \lambda_n=\sin^2\theta_1\,\sin^2\theta_2\dots \sin^2\theta_{n-k-1}}\nonumber\end{eqnarray} where $\theta_i \in [0,\ \pi/2],\,i=1,\dots,n-k-1$ are arbitrary angles. 
If $H$ is not positive definite,  the eigenvalues are parameterized by similar formulas,  as in {\bf Corollary 1}.
\end{Co}

 The Grassmann manifold, $Gr(k,n)$, is  defined as the set  of all $k$-dimension-onal subspaces of  ${\db C}^n$, and a main problem is to have a simple description of this variety. From what was said before,   evidently   there is a one-to-one correspondence between $k$-dimensional subspaces,  $k$-dimensional projections and a unitary operator like  $A(n,k)$. The first problem is to determine the number of independent parameters entering   $A(n,k)$ for this case.  Taking into account that  $k$- and $n-k$-dimensional subspaces  are  transformed into  $k$- and respectively $n-k$-dimensional subspaces  under the action of matrices from $U(k)\subset U(n)$ and respectively  $U(n-k)\subset U(n)$, $Gr(k,n)$ is viewed as the coset space

\begin{eqnarray}{ Gr(k,n) \cong \frac{U(n)}{U(k)\times U(n-k)} \cong \frac{Fl(n)}{Fl(k)\times Fl(n-k)}}
\end{eqnarray}
and by consequence the real dimension of the Grassmann manifold $ Gr(k,n)$ equals 

\begin{eqnarray}{d=n^2-k^2-(n-k)^2=2\,k(n-k)\label{dim}}
\end{eqnarray} 

Like the case of Stiefel manifolds, we 
  want first to obtain  a parameterization of $A(n,k)$ which is equivalent with finding a parameterization of Grassmannians.  With that end in view we rewrite relation (4.9) into the form
 \begin{eqnarray}
{M_n=A(n,k)\left(\begin{array}{cc}
B_{k}&0\\
0& I_{n-k}
\end{array}\right)\times
\left(\begin{array}{cc}
I_{k}&0\\
0& C_{n-k}
\end{array}\right)}
\end{eqnarray}
where  $M_n\in U(n)$ is an arbitrary matrix from $U(n)$ and  $B_{k}\in U(k)$ and respectively $ C_{n-k}\in U(n-k)$. Such a relation is always possible since any group operation is uniquely written as the product of an element in the subgroup with an element of the coset. From that relation we infer that the number of independent parameters entering $A(n,k)$ is the same as the dimension $d$ of the Grassmann manifold $Gr(k,n)$.

If we look at the relations  (\ref{matt}) or  (\ref{matttt}) and consider one of them for the case   $k=1$ we observe that  $B_n^0$
 is  equal to $A(n,1)$. Thus we obtain a matrix representation for
\begin{eqnarray}{ I_n -P=Gr(1,n)=A(n,1)\left(\begin{array}{cc}
1&0\\
0&0_{n-1}
\end{array}\right)A^*(n,1)} \end{eqnarray}
 i.e. the simplest Grassmannian, a result already known. The preceding relation can be written in  an equivalent form as 
\begin{eqnarray}{Gr(1,n)=v_1\cdot v_1^*}\end{eqnarray}
where $v_1$ is the vector that generates $B_n^0$, i.e. the vector $y_n$ from relation (\ref{Ve}).  
Thus $A(n,1)$ can be obtained from  (\ref{matt}) by taking  all the phases and angles entering $U(n-1)$ equal to zero, i.e. $U(n-1)=I_{n-1}$,  and here we give its explicit form

 \begin{eqnarray}{A(n,1)=\left(\begin{array}{cccc}
1&&&\\
&e^{i\alpha_1}&&\\
&&\dots&\\
&&&e^{i\alpha_{n-1}}
\end{array}\right)}\label{grass}\end{eqnarray}
 \begin{eqnarray}{
\left(\begin{array}{cccccc}
\cos\,a_1&-\sin\,a_1&0&0&\dots&0 \\
\sin\,a_1\,cos\,a_2&\cos\,a_1\,\cos\,a_2& -\sin\,a_2&0&\dots&0\\
\dots&\dots&\dots&\dots&\dots&0 \\
\dots&\dots&\dots&\dots&\dots&-\sin\,a_{n-1}\\
\sin\,a_1\dots \sin\,a_{n-1}&\dots&\dots&\dots&\dots&\cos\,a_{n-1}
\end{array}\right)\nonumber
}\end{eqnarray}
where we factored out the diagonal phase matrix.  This formula provides   an explicit  form of  the vectors $v_i$ that are  generated by {\bf Theorem 1}.

 In the following we show that 
 $ A(n,k) $ can be obtained in a similar way. Taking into account the form of the projection operator $ I_n-P$ and the dimensions $d_1$ and $d$ for Stiefel and Grassmann manifolds respectively, we infer that the Grassmann manifold is a special case of a Stiefel manifold. Our problem now is to find those constraints  which lead to the correct parameterization of $A(n,k)$ for Grassmannians.

 In order to view which are the constraints, let us consider the case 
 $k=2$. By taking into account that the first column of $A(n,2)$ coincides with the first column of the matrix (\ref{grass}),  we infer from (\ref{dim})  that the parameterization of the  second column is given in terms of  $2(n-3)$ new  real parameters. 
This means that we have to remove an angle and a phase from the vector $y_{n-1}$, equation (\ref{Ve}). A convenient choice is to take equal to zero the last angle and phase, i.e. $b_{n-2}=\beta_{n-2}=0$. This choice induces the following form for the matrix
$B_{n-1}^1$
 \begin{eqnarray}{B_{n-1}^1(b_{n-2}=\beta_{n-2}=0)=B_{n-1}^{1,1}=
\left(\begin{array}{ccc}
1&0&0\\
0&B_{n-2}&0\\
0&0&1
\end{array}\right)}\end{eqnarray}
where $B_{n-2}$ is generated as in {\bf Theorem 1} by the vector
 \begin{eqnarray}{
y_{n-2}^{'}=(\cos\,b_1,e^{i\beta_1}\sin\,b_1\,\cos\,b_2,\dots,e^{i\beta_{n-3}}\sin\,b_1\dots \sin\,b_{n-3})^t}\end{eqnarray}
 It is easily seen that this structure preserves from $k\rightarrow k+1$ such that 
\begin{eqnarray}{B_{n-k}^{k,k}=\left(\begin{array}{ccc}
I_k&0&0\\
0&B_{n-2k}&0\\
0&0&I_k
\end{array}\right),\qquad k=1,\dots,[\frac{n}{2}] }\end{eqnarray}
where $[a]$ denotes the integer part of $a$, and $B_{n-2k},\,  k=1,\dots,[\frac{n}{2}], $ are generated by the vectors
\begin{eqnarray}{\begin{array}{l}
w_1=(\cos\,a_1,e^{i\alpha_1}\,\sin\,a_1\,\cos\,a_2,\dots,,e^{i\alpha_{n-1}}\,\sin\,a_1\dots \sin\,a_{n-1})^t\\
\\
w_{2}=(\cos\,b_1,e^{i\beta_1}\sin\,b_1\,\cos\,b_2,\dots,e^{i\beta_{n-3}}\sin\,b_1\dots \sin\,b_{n-3})^t\\
..........................................................................................................\\
 w_p=(\cos\,l_1,e^{i\varphi_1}\sin\,l_1\,\cos\,l_2,\dots,e^{i\varphi_{n-(2p-1)}}\sin\,l_1\dots \sin\,l_{n-(2p-1)})^t\\
...........................................................................................................\\
w_{[n/2]}=(\cos\,z_1,e^{i\omega_1}\,sin\,z_1\,\cos\,z_2,e^{i\omega_2}\,\sin\,z_1\,\sin\,z_2 )^t,\,\, {\rm for}\,\, n \,\,{\rm odd}\\
\\
w_{[n/2]}=(\cos\,z_1,e^{i\omega_1}\,\sin\,z_1 )^t,\,\, {\rm for}\,\, n \,\,{\rm even}
\end{array}}
\end{eqnarray}

With the above  notation the following holds
\setcounter{Le}{2}
\begin{Le}
The unitary matrix $A(n,k)\in U(n)$ entering the matrix representation of Grassmann manifold has the form
\begin{eqnarray}{ A(n,k)=B_n^0\cdot B_{n-1}^{1,1}\dots B_{n-k+1}^{k-1,k-1},\,\, k=1,\dots,[\frac{n}{2}]}
\end{eqnarray}
 and the matrix representation of the projection onto the $k$-dimensional subspace of ${\db C}^n$   is
 \begin{eqnarray}{I_n-D_{T^*}=Gr(k,n)= A(n,k)\left(\begin{array}{cc}
I_k&0\\
0&0
\end{array}\right) A(n,k)^*}\end{eqnarray}
An equivalent description is given by
\begin{eqnarray}{
Gr(k,n)=\sum_1^k\,u_i\cdot u_i^*}\end{eqnarray}
where $ u_i,\, i=1,\dots,k$, are the first $k$ column vectors of matrix {\rm (4.19 )}.

\end{Le}

\setcounter{Co}{2}
\begin{Co}

If $H$ is a Hermitian operator that  has only two eigenvalues whose  multiplicities are $k$ and $n-k$ then
 \begin{eqnarray}{H=\lambda_1\,\sum_{i=1}^k\,u_i u_i^* + \lambda_2\,\sum_{i=k+1}^n \,u_i u_i^*}\end{eqnarray}
where $u_i,\,i=1,\dots,n$, are the column vectors of the matrix $A(n,k)$, equation {\rm(4.19)}. If $H$ is a density matrix operator, then $\lambda_1=\cos^2\theta$ and $\lambda_2=\sin^2\theta$, where $\theta\in [0,\pi/2]$ is an arbitrary angle. With this choice, formula {\rm(4.22)} describes all the density matrices that are   solutions of the equation {\rm(1.2)}.

When $H$ is not positive definite, and $Tr\,H=h$, the parameterization of eigenvalues is:  $\lambda_1=h\,\cosh^2\theta$ and $\lambda_2=-h\, \sinh^2\theta$, if $\lambda_1 > |\lambda_2|$, and  $\lambda_1=|h|  \sinh^2\theta$ and $\lambda_2=-|h|\cosh^2\theta$, if $h < 0$.
           .
\end{Co}
\end{newtheorem}

In this way we succeeded to find an explicit  parameterization for the finite-dimensional Hermitian operators, in    three  cases:  the generic case, when all the eigenvalues are different, an intermediate case when the first $k$ eigenvalues are degenerated, the other eigenvalues being simple, and the most ``degenerated'' case, when there exist only two (different) eigenvalues. 
Matrix realization of other flag manifolds,  as e.g. 
\begin{eqnarray}{Fl(k_1,k_2,\dots,k_l,;n)\cong\frac{Fl(n)}{Fl(k_1)\times Fl(k_2)\times\dots\times Fl(k_l) }}\end{eqnarray} with $\sum_i^l\,k_i\leq n$, that describe other spectral multiplicities,  can be found in a similar way.
Assembling together all the previous information  we have
\setcounter{Co}{2}
\begin{Co}
The flag matrix {\rm (\ref{matt})} encodes all the possible spectral decompositions of  finite-dimensional  Hermitian operators, acting on ${\db C}^n$, i.e. any other particular unitary matrix $A(n,k)$ that describes a given spectral multiplicity of an Hermitian operator can be obtained from it by properly restricting the number of parameters entering {\rm(\ref{matt})}.
\end{Co}
\vskip2mm
{\it Example}.
We consider  that a simple example of a  matrix realization of a Grassmann manifold will help, and  we consider the case  $n=8$ and $k=4$; then $A(8,4)$ has the form:
%\newpage

 \begin{eqnarray}{A(8,4)=}\nonumber\end{eqnarray}
 \begin{eqnarray}{
 \left(\begin{array}{cccccc}
\cos a_1&-\sin a_1&0&0&\dots&0 \\
e^{i\alpha_1}\sin a_1\,\cos a_2&e^{i\alpha_1}\cos a_1\,\cos a_2& -e^{i\alpha_1}\sin a_1&0&\dots&0\\
\dots&\dots&\dots&\dots&\dots&0 \\
\dots&\dots&\dots&\dots&\dots&-e^{i\alpha_{6}}\sin a_{7}\\
e^{i\alpha_{7}}\sin a_1\dots \sin a_{7}&\dots&\dots&\dots&\dots&e^{i\alpha_{7}}\cos a_{7}
\end{array}\right)\times}\nonumber\end{eqnarray}
\begin{eqnarray}{
\left(\begin{array}{cccccc}
1&0&0&0&\dots&0 \\
0&-\cos b_1&\sin b_1&0&\dots&0\\
0&e^{i\beta_{1}}\sin b_1\, \cos b_2&e^{i\beta_{1}}\cos b_1 \cos b_2   &-e^{i\beta_{1}}\sin b_2&\dots&0 \\
\dots&\dots&\dots&\dots&\dots&-e^{i\beta_{4}}\sin b_{5}\\
0&e^{i\beta_{5}}\sin b_1\dots \sin b_5&\dots&\dots&\dots&e^{i\beta_{5}}\cos b_{5}\\
0&\dots&\dots&\dots&\dots&1
\end{array}\right)\nonumber
}\nonumber\end{eqnarray}
\begin{eqnarray}{\left(\begin{array}{cccccc}
I_2&0&0&0&0&0\\
0&-\cos c_1& \sin c_1&0&0&0\\
0&e^{i\gamma_{1}}\sin c_1 \cos c_2&e^{i\gamma_{1}}\cos c_1 \cos c_2&-e^{i\gamma_{1}}\sin c_2&0&0\\
0&e^{i\gamma_{2}}\sin c_1 \sin c_2 \cos c_3&e^{i\gamma_{2}}\cos c_1 \sin c_2 \cos c_3&e^{i\gamma_{2}}\cos c_2 \cos c_3&-e^{i\gamma_{2}}\sin c_3&0\\
0&e^{i\gamma_{3}}\,\sin c_1 \sin c_2 \sin c_3&e^{i\gamma_{3}}\cos c_1 \sin c_2 \sin c_3&e^{i\gamma_{3}}\,\sin c_2 \cos c_3&e^{i\gamma_{3}}\cos c_3&0\\
0&0&0&0&0&I_2
\end{array}\right)\times}\nonumber
\end{eqnarray}
\begin{eqnarray}{
\left(\begin{array}{cccc}
I_3&0&0&0\\
0&-\cos d_1&\sin d_1&0\\
0&e^{i\delta_{1}}\sin d_1&e^{i\delta_{1}}\cos d_1&0\\
0&0&0&I_3
\end{array}\right)
}\end{eqnarray}
and the matrix representation of $Gr(k,8)$ is
 \begin{eqnarray}{Gr(k,8)=\sum_{i=1}^{i=k} u_i\,u_i^*,\quad{\rm for}\,\, k=1,\dots,4}\end{eqnarray}
where $u_i,\,i=1,\dots,k$ are the first $k$ columns of matrix (4.24). In the same time formula (4.24) provides us all the Grassmannians $Gr(k,n)$ for $n \leq 7$. For example $A(k,7)$ is obtained from (4.24) by putting $a_7=b_5=c_3=d_1=\alpha_7=\beta_5=\gamma_3=\delta_1=0$ and by deleting the last row and   column, and so on.

%%%%%%%%%%%%%%%%%%%%%%%%%%%%%%%%%%%%%%%%%%%%%%%%%%%%%%
\section{Conclusion}
%%%%%%%%%%%%%%%%%%%%%%%%%%%%%%%%%%%%%%%%%%%%%%%%%%%%%%
In this paper we have obtained a  constructive parameterization of all finite-level Hermitian operators. Further, this construction is recursive: if we have a parameterization of $\got{M}_n$, $\got{M}_{n+2}$ is obtained by embedding  $\got{M}_n$ into $\got{M}_{n+2}$, and by  multiplication at left by an appropriate matrix of the form (\ref{grass}), with $n\rightarrow n+2$.   We have shown that the unitary matrices which diagonalize the Hermitian operators are subsets of the flag unitary matrix $\got{M}_n$ whose explicit form is given by formula (\ref{matt}). When the eigenvalues are simple, the generic form of the  unitary operators is $\got{M}_n$; when there is a $k$-fold degeneracy, the corresponding unitary matrices are in the set $St(k,n)$. If there is a maximum degeneracy, i.e. only two eigenvalues, with multiplicities $k$ and, respectively $n-k$,  are distinct, the unitaries are in the Grassmannian $G(k,n)$. Our explicit construction was done only for the $n$-dimensional unitary group $U(n)$, but it is evident that the same approach works in the case of any compact group. For example, taking zero all the phases entering $U(n)$ and doing similar calculations, one gets results for $SO(n)$, and so on.

We consider that the above parameterization will be useful for doing calculation, especially for problems where one has to make an optimization over a set of parameters, e.g. for the characterization of entanglement. 
 
Taking  also into account our previous results \cite{Di2}, which state that any unitary matrix entering  $\got{M}_n$ is an ordered product of $n-1$ diagonal phase matrices and $n(n-1)/2$ two-dimensional rotations, one can make use of the device designed by Reck {\it et al} \cite{ RZBB} to give an operational meaning in the real world to any finite-level Hermitian operator. This means that all the Hermitian matrices could be experimentally implemented and, by consequence, could be measured. Such multi-states devices will find applications in quantum information processing and quantum computation.

In the same time, we obtained a new and simple analytic representation of Stiefel and Grassmann manifolds,  representation that is essentially contained in a unitary matrix $A(n,k)$, which  can be easy stored into a computer, and problems similar to  those encountered in \cite{EAS}, or \cite{ZT} will be  easier to solve. 

Our results show also that there are two new symmetric spaces $\got{M}_n$, and $St(k,n)$, defined as in {\rm(\ref{St})}, on which one can implement a symplectic structure, and an interesting problem would be the finding of the relationship between a generic Hermitian operator on these symmetric spaces and the  Laplace-Beltrami operator on the corresponding  manifold.  Some of these problems will be considered elsewhere.

\section*{Acknowledgments} The author wants to thank the anonymous referee of his paper \cite{Di2} who recommended the following: \begin{quote} [\dots] I recommend that the factorization theorem be stated clearly in the early part of the paper and that it be followed by a construction algorithm. [\dots] \end{quote} Making explicitly this algorithm I realized that it can be used to obtain novel matrix realizations for compact symmetric manifolds.

\end{document}